\documentclass[]{aa}

\def\rmit#1{{\it #1}}              
\def\specchar#1{{\sc #1}}

\def\CaIIH{\mbox{Ca\,\specchar{ii}\,\,H}}       
\def\CaII{\mbox{Ca\,\specchar{ii}}}

\def\eg{\rmit{e.g.}}

\usepackage{booktabs}
\usepackage{graphicx}
\usepackage{multirow}
\usepackage{color}
\usepackage[flushleft]{threeparttable}

\usepackage{array}
\newcolumntype{?}{@{\vrule width 2pt}}

\usepackage{hyperref}
\hypersetup{
    final=true,
    pageanchor=true,
    colorlinks=true,
    breaklinks=true,
    linkcolor=blue,
    citecolor=blue,
    urlcolor=blue,
    pdfpagemode=UseNone,
    pdftitle={Inversions of synthetic umbral flashes: selection of wavelength sampling},
    pdfauthor={T. Felipe et al.},
    pdfsubject={Solar Physics},
    pdfkeywords={Methods: numerical -- Sun: chromosphere -- sunspots -- Sun: oscillations -- Techniques: polarimetric}}

\titlerunning{Selection of wavelength sampling from synthetic umbral flashes}   

\begin{document}



\title{Inversions of synthetic umbral flashes: a selection of wavelength sampling}

\author{T. Felipe\inst{\ref{inst1},\ref{inst2}}
\and S. Esteban Pozuelo\inst{\ref{inst1},\ref{inst2}}
}


\institute{Instituto de Astrof\'{\i}sica de Canarias, 38205, C/ V\'{\i}a L{\'a}ctea, s/n, La Laguna, Tenerife, Spain\label{inst1}
\and 
Departamento de Astrof\'{\i}sica, Universidad de La Laguna, 38205, La Laguna, Tenerife, Spain\label{inst2} 
}

\abstract
{Imaging spectrographs are popular instruments used to obtain solar data. They record quasi-monochromatic images at selected wavelength positions. By scanning the spectral range of the line, it is possible to obtain bidimensional maps of the field-of-view with a moderate spectral resolution.} 
{In this work, we evaluate the quality of spectropolarimetric inversions obtained from various wavelength samplings during umbral flashes.}
{We computed numerical simulations of nonlinear wave propagation in a sunspot and constructed synthetic Stokes profiles in the \CaII\ 8542 \AA\ line during an umbral flash using the NLTE code NICOLE. The spectral resolution of the Stokes profiles was downgraded to various cases with differences in the wavelength coverage. A large set of wavelength samplings was analyzed and the performance of the inversions was evaluated by comparing the inferred chromospheric temperature, velocity, and magnetic field with the actual values at the chromosphere of the numerical simulation.}
{The errors in the inverted results depend to a large extent on the location of the wavelength points across the profile of the line. The inferred magnetic field improves with the increase of the spectral resolution. In the case of velocity and temperature, low spectral resolution data produce a match of the inverted atmospheres with the actual values comparable to wavelength samplings with finer resolution, while providing a higher temporal cadence in the data acquisition.}
{We validated the NLTE inversions of spectropolarimetric data from the \CaII\ 8542 \AA\ during umbral flashes, during which the atmosphere undergoes sudden dramatic changes due to the propagation of a shock wave. Our results favor the use of fine spectral resolution for analyses that focus on the inference of the magnetic field, whereas the estimation of temperature and velocity fluctuations can be performed with lower spectral resolution.}


\maketitle


\section{Introduction}

Umbral flashes (UFs) are sudden intensity enhancements that take place in the core of the \CaII\ lines. They were first detected by \citet{Beckers+Tallant1969}, who reported a periodicity around three minutes and soon after other works confirmed their properties and their origin \citep{Wittmann1969, Havnes1970}. It has been well established that UFs are produced by the propagation of magneto-acoustic waves along the umbral field lines, which come directly from the photosphere. The amplitude of these waves in the three-minute band increases with height due to the drop of the density and they go on to develop into shocks at the chromosphere \citep{Centeno+etal2006, Felipe+etal2010b}.   

Spectropolarimetric observations of UFs have been common since the beginning of our century. Using the \CaII\ infrared triplet lines, \citet{SocasNavarro+etal2000a, SocasNavarro+etal2000b} detected the periodic occurrence of anomalous circular polarization profiles associated with UFs. They were interpreted as a two-component atmosphere inside the resolution element, where one of the atmospheres is magnetized and slightly downward-flowing and the other component exhibits a strong upward velocity around 10 km s$^{-1}$. The \CaII\ 8542 \AA\ line has been one of the most popular spectral lines for the study of UFs using polarimetry. Various works have analyzed this line using observations acquired with the CRISP instrument \citep{Scharmer+etal2008} attached to the 1-m Swedish Solar Telescope \citep[SST,][]{Scharmer+etal2003}. They have confirmed the fine structure of UFs and measured the fluctuations in temperature, velocity, and (in some cases) magnetic field \citep{delaCruz-Rodriguez+etal2013, Henriques+etal2017, Joshi+delaCruzRodriguez2018}. 

The use of instruments based on the Fabry-P\'erot interferometer system, such as CRISP, calls for the search for a compromise between the spectral resolution and the temporal cadence. These instruments perform the scan of the spectral line by successively acquiring bidimensional images at selected wavelength points across the line profile. There are significant differences among the wavelength samplings employed by various works studying sunspots in the \CaII\ 8542 \AA\ line, including differences in the spectral resolution around the core of the line and the wings coverage. The spectroscopic (without polarimetry) \CaII\ 8542 \AA\ observations from \citet{Houston+etal2018} used a fine resolution, including 47 non-equidistant points in the spectral range $\Delta\lambda=\pm1300$ m\AA, with $\Delta\lambda$ defined as the wavelength shift with respect to the core of the \CaII\ 8542 \AA\ line. However, the acquisition of full spectropolarimetry requires further measurements and given the temporal scales of chromospheric fluctuations, observers generally cannot afford this fine resolution when gathering polarimetric data. Some spectropolarimetric observations of sunspots have employed a moderate spectral resolution around the core of the \CaII\ 8542 \AA\ line \citep[$\delta\lambda\approx 100$ m\AA, e.g.,][]{WedemeyerBohm2010, EstebanPozuelo+etal2019}. Those observations focusing on UFs generally employ a spectral resolution around 70-75 m\AA\ \citep{delaCruz-Rodriguez+etal2013, Henriques+etal2017, Joshi+delaCruzRodriguez2018}. In all cases, it is a common approach to employ a lower resolution at the wings of the line.

The aim of this work is to evaluate the performance of the inversions of spectropolarimetric data from the \CaII\ 8542 \AA\ line using various wavelength samplings. We focus on the analysis of UFs since it is over the duration of such events that the umbra experiences sudden changes in the chromospheric atmosphere. We employed numerical simulations and the computation of synthetic Stokes profiles. Previous works have also simulated the profiles of \CaII\ lines during UFs. \citet{Bard+Carlsson2010} performed hydrodynamic simulations to investigate the formation of UFs in the intensity profiles of the \CaIIH\ line. \citet{Felipe+etal2014b} synthesized the four Stokes parameters in the \CaII\ infrared triplet from magnetohydrodynamic (MHD) simulations of nonlinear waves driven by fluctuations measured from photospheric observations \citep{Felipe+etal2011}. More recently, \citet{Felipe+etal2018a} constructed synthetic Stokes profiles of the \CaII\ 8542 \AA\ line that take into account the time required by imaging spectropolarimeters to acquire the various wavelength points across the profile of the spectral line. Here we follow the same strategy from \citet{Felipe+etal2018a} but we test the results obtained from the inversions of line profiles with the differences in their wavelength sampling.

The organization of the paper is as follows: in Sect. \ref{sect:methods} we describe the numerical simulations and code used for the synthesis and inversion of the Stokes profiles; in Sect. \ref{sect:sampling} we present the set of wavelength samplings explored in this work; in Sect. \ref{sect:inversions} we briefly discuss the results of the inversions for one of the cases; and finally Sects. \ref{sect:evaluation} and \ref{sect:conclusions} present the results and conclusions, respectively.


\section{Numerical methods}
\label{sect:methods}

The propagation of slow magnetoacoustic waves from the solar interior to the chromosphere was simulated using the code MANCHA \citep{Khomenko+Collados2006, Felipe+etal2010a}. The code solves non-linear ideal MHD equations, computing the response of a magnetohydrostatic sunspot atmosphere \citep{Khomenko+Collados2008,Przybylski+etal2015} to perturbations initially located below the solar atmosphere. The umbral photospheric magnetic field strength of the sunspot model is 2300 G and its Wilson depression is 450 km. The simulation has been performed using the 2.5D approximation, that is, only the vertical and one of the horizontal dimensions are included in the calculation but the three components of the vectors are maintained. In the horizontal direction, the computational domain covers 140 Mm, with the axis of the sunspot located at the middle position and using a horizontal spatial step of 0.2 Mm (comparable to the spatial resolution achieved by current solar telescopes). The vertical direction was sampled using a grid with variable step size based on the sound speed, so the acoustic travel time between neighboring cells is constant. A larger vertical spatial step was used in regions with a higher sound speed, except for layers above the temperature minimum, where the minimum resolution of $\Delta z=15$ km is employed. The coarser vertical resolution ($\Delta z=44$ km) is found at the bottom boundary. In this work, we focus on the analysis of a single UF event that takes places near the center of the umbra a few minutes after the beginning of the simulation. We analyze the temporal variation of the atmosphere at one of the vertical columns of the MHD simulation. The output of the simulation, including temperature, velocity, pressure, density, and magnetic field, was saved every 0.120 s. More details on the numerical simulation can be found in Appendix A from \citet{Felipe+etal2018a}.

The non-local thermodynamic equilibrium (NLTE) code NICOLE \citep{SocasNavarro+etal2015} was employed for synthesizing and inverting the four Stokes profiles of the \CaII\ 8542 \AA\ line prior, during, and after the UF event. For the synthesis, the atmospheric models were provided to NICOLE as they resulted from the simulation, with the vertical stratification given in geometrical scale. An artificial microturbulent velocity of 3 km s$^{-1}$ was included to compensate the absence of small scale random motions in the simulations and produce \CaII\ 8542 \AA\ intensity profiles with a width comparable to those reported in actual observations \citep{delaCruz-Rodriguez+etal2012}. The original synthetic profiles were computed with a  spectral sampling of 5 m\AA. They were then convolved with a Gaussian spectral filter with a full width half maximum of 100 m\AA\ at the wavelength positions selected for the spectra sampling (for more, see the next section).   

Imaging spectrographs construct the line profiles by sequentially scanning bidimensional images of the field of view at different wavelengths, sampling the whole profile. This way, each profile is composed of images obtained at different time steps. We constructed synthetic profiles that imitate the acquisition strategy of imaging 
spectrographs. More specifically, we mimicked the timing of the scanning obtained from spectropolarimetric observations acquired with CRISP at the SST. Its cameras obtain images at 35 frames per second. Taking into account that full spectropolarimetry requires four images per wavelength position, each accumulation requires 0.114 s. The total integration time of each wavelength is given by the number of accumulations selected for that position. We constructed the synthetic profiles assuming that each time step from the numerical simulation corresponds to one accumulation. Note that the time required per accumulation in actual observations (0.114 s) is slightly shorter than the temporal cadence chosen for the output of the simulation (0.120 s). This additional time partially compensates for the frames which are discarded when the Fabry-P\'erot is changing the wavelength of the transmission peak (about two frames per wavelength change).

For all the cases explored in this paper, the line profile was scanned starting at the blue extreme and increasing the wavelength in subsequent measurements (at successive time steps). The Stokes signal at each wavelength was computed as the average of the value of the Stokes parameter for the number of accumulations employed (generally, four accumulations at the wings and seven accumulations around the core of the line), as obtained from the NICOLE synthesis of successive time steps from the numerical simulation. Random noise was added to each independent accumulation. After averaging, the noise is of the order of $1\times10^{-3}$ in units of continuum intensity, similar to that obtained in UF observations \citep{delaCruz-Rodriguez+etal2013}. The analyses were repeated with a noise of $2\times10^{-4}$ and no significant differences were found.

The synthetic profiles were then inverted using NICOLE. In the inversion mode, a user-supplied initial guess atmosphere \citep[in our case, the FAL-C mode from][]{Fontenla+etal1993} is iteratively modified until agreement is found between the input profiles and those generated by the resultant atmosphere. The inversions were performed using two cycles: the first one with a modest number of nodes (four nodes in temperature, two nodes in velocity, and one node in vertical magnetic field) and the second one using a greater number of nodes (eight in temperature, five in velocity, two in vertical magnetic field, and one in the other two components of the magnetic field and the microturbulence). For the inversion, we used a weight of 70\%
of that of Stokes I for Stokes $V$ , along with a 50\% weight for Stokes Q and U. Since the atmospheres analyzed in this work correspond to the central part of the umbra, with a mostly vertical magnetic field, the signals in Stokes Q and U are very weak. Nevertheless, we included those signals in order to constrain the results from the inversion to atmospheric models that generate weak linear polarization signals.

\section{Spectral sampling}
\label{sect:sampling}

\begin{figure*}[!ht] 
 \centering
 \includegraphics[width=18cm]{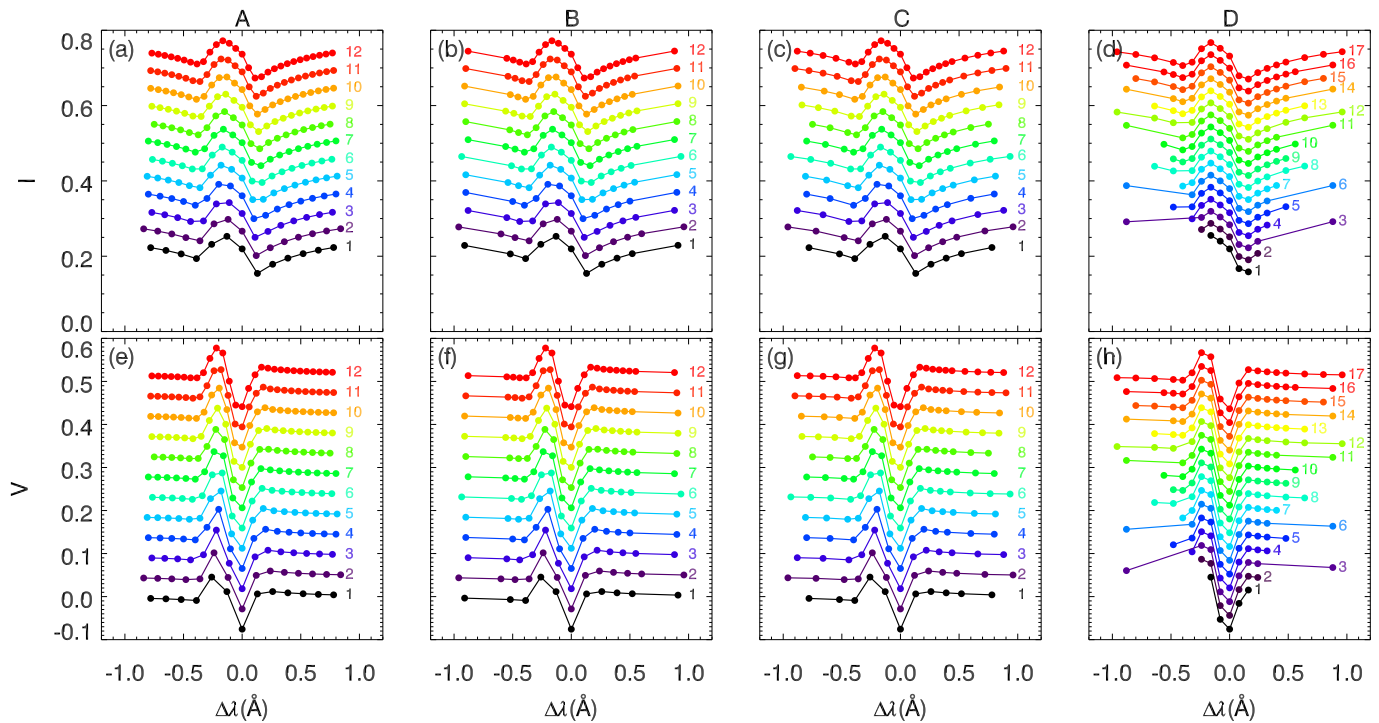}
  \caption{Wavelength samplings of the \CaII\ 8542 \AA\ line in Stokes $I$ (top panels) and Stokes $V$ (bottom panels) during a simulated UF. Each column corresponds to a set of wavelength samplings constructed according to a different criteria. The profiles have been shifted in the vertical axis to support the visualization.}
    
  \label{fig:sampling}
\end{figure*}

\begin{table*}
\begin{center}
\caption[]{\label{table:samplings}
          {Summary of criteria used for selecting wavelength samplings.}}
\begin{tabular}{lcccc}
\hline\noalign{\smallskip}
Criterion                               &       Core            &       Wing                                         & Sampling                              &  RMSE                   \\
                                        &       resolution      &        resolution                                              & representation                        &                         \\
\hline\noalign{\smallskip}      
A                                       &       $\delta\lambda$         &         $\delta\lambda$                                 & Fig. \ref{fig:sampling}, first column    & Fig. \ref{fig:RMSE}, blue line        \\      
B                                       &       $\delta\lambda$         &         2 points, $\Delta\lambda\approx\pm0.9$ \AA\     & Fig. \ref{fig:sampling}, second column   & Fig. \ref{fig:RMSE}, black line       \\      
C                                       &       $\delta\lambda$         &         2$\delta\lambda$                                & Fig. \ref{fig:sampling}, third column    & Fig. \ref{fig:RMSE}, red line \\      
D                                       &       $\delta\lambda=80$ m\AA\ &       Variable                                        & Fig. \ref{fig:sampling}, fourth column   & Fig. \ref{fig:RMSE_dl80}      \\      
\hline

\end{tabular}

\begin{tablenotes}
\small
 \item {Each row (A, B, C, D) corresponds to a set of wavelength samplings constructed following a different criterion. The second column indicates the spectral resolution in the core of the line, the third column is the spectral resolution at the wings, the fourth column points to the figure and panels where the samplings are illustrated, and the last column indicates the figure where the measured RMSE are plotted. Twelve wavelength samplings with different values of the spectral resolution in the core ($\delta\lambda=55, 60, 65, 70, 75, 80, 85, 90, 100, 110, 120, 130$ m\AA) are analyzed for criteria A, B, and C. In the case of criterion D, all the 17 wavelength samplings use the same spectral resolution at the core $\delta\lambda=80$ m\AA. They differ in the wavelength region sampled around the core, the resolution at the wings, and the maximum wavelength included.  }
\end{tablenotes} 
  
\end{center}
\end{table*}

The goal of this work is the identification of the wavelength coverages that produce the best results for the analysis of the commonly-used \CaII\ 8542 \AA\ line during UFs. We are not only interested in a good match between the results from the inversion and the actual atmospheric data, but also in obtaining a fast temporal cadence that allows for the study of the sudden changes that take place in the umbral chromosphere during the propagation of shock waves. 

We performed a parametric study by testing many different wavelength configurations. The columns in Fig. \ref{fig:sampling} illustrate each set of wavelength samplings constructed under a certain criteria (see Table \ref{table:samplings}). The top and bottom panels show the Stokes $I$ and $V$ profiles, respectively, produced by the same atmospheric model at a randomly selected time step during the development of the UF.

For the criterion A (first column from Fig. \ref{fig:sampling}), the whole profile between approximately $\Delta\lambda=-0.8$ \AA\ and $\Delta\lambda=0.8$ \AA\ is sampled with a constant wavelength step ($\delta\lambda$), with one of the points located at $\Delta\lambda=0$. Twelve different samplings were selected, from a fine spectral resolution ($\delta\lambda=55$ m\AA, profile \#12 in panels a and e, this value is close to the Nyquist frequency of the CRISP instrument at 8542 \AA) to a coarse resolution ($\delta\lambda=130$ m\AA, profile \#1 in panels a and e). Since the wavelength positions are multiples of the chosen $\delta\lambda$, the maximum $|\Delta\lambda|$ covered by the sampling is different for each case. We chose to extend the scanning to the wavelength closer to $|\Delta\lambda|=0.8$. 

For the wavelength samplings constructed according to criterion B (second column of Fig. \ref{fig:sampling}), a constant wavelength step $\delta\lambda$ is employed around the core of the line, between approximately $\Delta\lambda=-0.5$ \AA\ and $\Delta\lambda=0.5$ \AA, and two additional wavelength points are added at the blue and red wings. These wing wavelengths are placed at the multiple of $\delta\lambda$ closer to $\pm0.9$ \AA. Similarly to the cases illustrated in the first column, we also defined twelve independent wavelength samplings following this criterion, from a fine resolution in the core of the line ($\delta\lambda=55$ m\AA, profile \#12 in panels b and f) to a coarse resolution ($\delta\lambda=130$ m\AA, profile \#1 in panels b and f).

The samplings from criterion C (third column of Fig. \ref{fig:sampling}) were constructed by imposing a constant $\delta\lambda$ around the core of the line (approximately in the wavelength range between $\Delta\lambda=-0.45$ \AA\ and $\Delta\lambda=0.45$ \AA) and a wavelength step of $2\delta\lambda$ in the wings, between about $\Delta\lambda=\pm0.45$ and $\Delta\lambda=\pm0.85$. 

In the last set of wavelength samplings (criterion D, fourth column from Fig. \ref{fig:sampling}), all of them employ a minimum wavelength resolution of $\delta\lambda=80$ m\AA, but they differ in the wavelength region covered around the core, the number of wavelength points and resolution in the wings, and the maximum wavelength included. The profiles are ordered from shorter scanning time to longer scanning time from bottom to top, that is, cases labeled with a higher number require more time to complete the scan of the whole profile.

\section{Inversions}
\label{sect:inversions}

\begin{figure*}[!ht] 
 \centering
 \includegraphics[width=18cm]{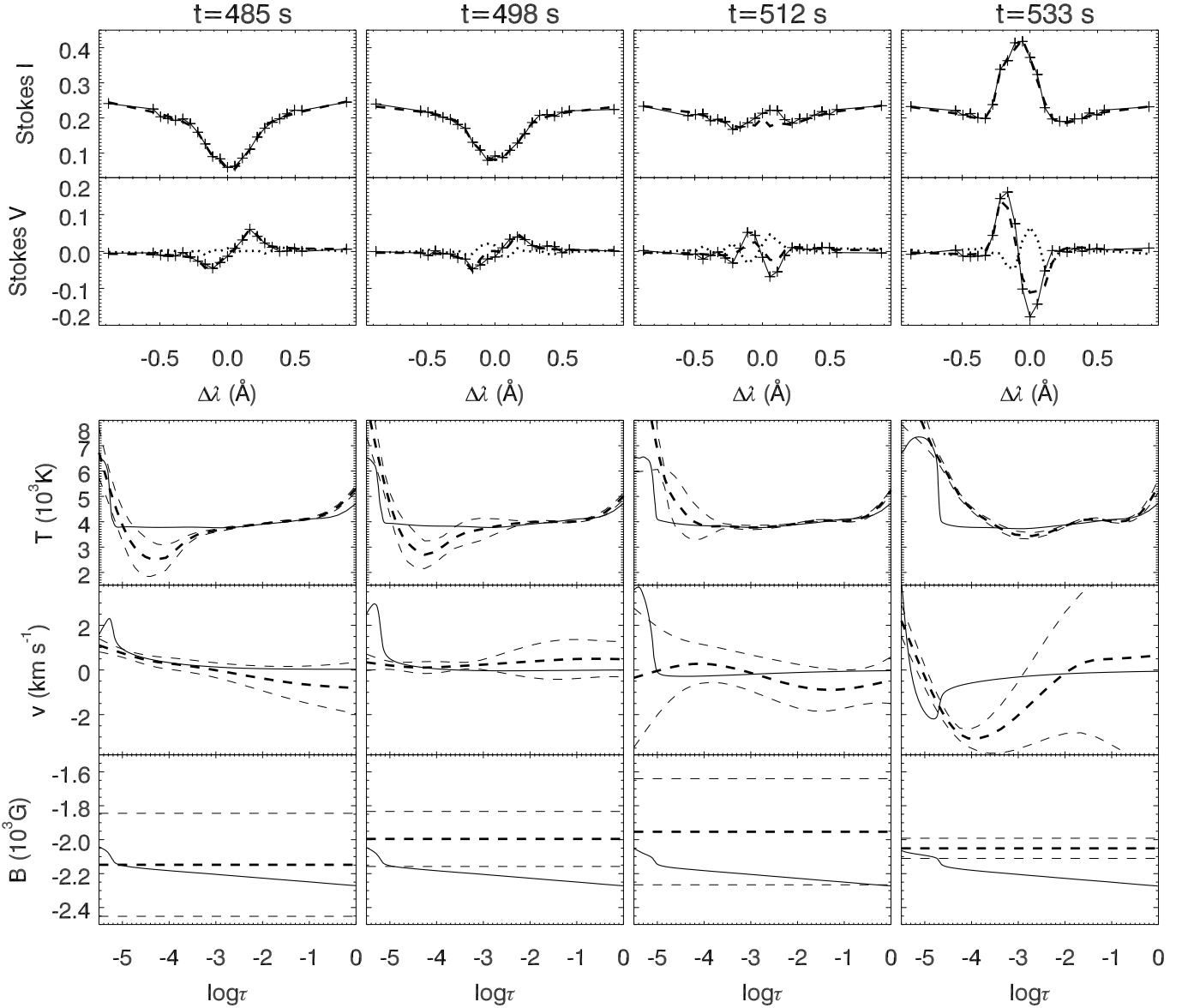}
  \caption{Inversions of four time steps during the development of a synthetic UF using a wavelength sampling of $\delta\lambda=55$ m\AA\ between $\Delta\lambda=\pm0.55$ \AA\ in addition to points at $\Delta\lambda=\pm0.88$ \AA\ (case \#12 from criterion B, second column in Fig. \ref{fig:sampling}). Top two rows: original (solid lines with crosses) and inverted (dashed lines) Stokes $I$ and $V$ profiles. Second row: the dotted lines illustrate the residuals from Stokes V. Bottom three rows: comparison between stratification of the simulated atmospheres (solid lines) and those inferred from the inversion of the synthetic Stokes profiles (dashed lines). Thick dashed lines correspond to the best inversion (cases with lowest $\chi^2$) and dashed lines illustrate the uncertainties.}    
  \label{fig:inversion}
\end{figure*}

The top panels of Fig. \ref{fig:inversion} illustrate the synthetic Stokes $I$ and $V$ profiles obtained at four time steps during the UF and their comparison with the profiles retrieved after the inversion. The bottom panels of the same figure show the actual temperature, velocity, and vertical magnetic field of the atmosphere, as given by the numerical simulation, and the atmospheric stratification inferred from the inversion of the synthetic Stokes profiles. The inversions have been performed every ten time steps from the output of the numerical simulation (that is, with a cadence of 1.2 s) and for all the wavelength samplings illustrated in Fig. \ref{fig:sampling}. In order to illustrate the performance of the inversion, we chose to show as an example the results of the analyses of four time steps for the profiles with a wavelength sampling of $\delta\lambda=55$ m\AA\ between $\Delta\lambda=\pm0.55$ \AA\ and points at $\Delta\lambda=\pm0.88$ (corresponding to case \#12 from criterion B, second column of Fig. \ref{fig:sampling}).

The standard approach for performing inversions with NICOLE consists of repeating each inversion several times using randomized initializations and selecting the one which gives a minimum in the merit function $\chi^2$. This way, the chance that the inversion settles into a local minimum is reduced. The second cycle of all the inversions computed in this work was repeated 30 times and the solution with the lowest $\chi^2$ was selected. The uncertainties shown in the bottom panels of Fig. \ref{fig:inversion} (thin dashed lines) are estimated as the standard deviation over five independent inversions following the aforementioned process, each including 30 inversions.

Figure \ref{fig:inversion} shows that the inversion captures the main features of the Stokes profiles of the \CaII\ 8542 \AA\ line produced by the UF and their temporal variation reasonably well. The intensity core, which appears in absorption for a quiescent atmosphere (two first columns), is progressively reversed until a prominent blueshifted emission core develops. During this core reversal, the Stokes $V$ profile exhibits a change in the polarity. 

The bottom panels of Fig. \ref{fig:inversion} illustrate a comparison between the inferred quantities and the actual atmospheric values (extracted directly from the simulation) as a function of optical depth. The temperature of the umbral model shows an approximately constant value with a sharp enhancement at the high chromosphere. As the shock wave reaches the chromosphere, the height of this temperature increase is shifted to lower layers. The inversion reproduces this change in the height of the temperature enhancement, although the inferred temperature gradient is more subtle. In order to interpret this mismatch, we have to take into account that the inversion code cannot determine variations in layers thinner than the photon mean free path, since the radiation would travel through them unchanged.

At the time steps illustrated in Fig. \ref{fig:inversion}, the chromospheric line-of-sight (LOS) velocity (around $\log\tau=-5$) exhibits a sudden change from a positive value (redshift, downflow) to a negative velocity (blueshift, upflow) at the last time step. This rapid variation is also captured by the inversion code. Interestingly, the velocity uncertainties at deeper layers are larger. This is expected from the selection of wavelength points employed in this case since it uses a good wavelength sampling at the core (with a chromospheric formation height) but a poor sampling at the wings of the \CaII\ 8542 \AA\ line. This wavelength sampling cannot adequately constrain the photospheric layers.

The simulated vertical magnetic field exhibits very small changes during the arrival of the shock wave. On the contrary, the values inferred from the inversion can exhibit fluctuations up to 200 G.

\section{Evaluation of the inferred chromosphere}
\label{sect:evaluation}

\begin{figure}[!ht] 
 \centering
 \includegraphics[width=9cm]{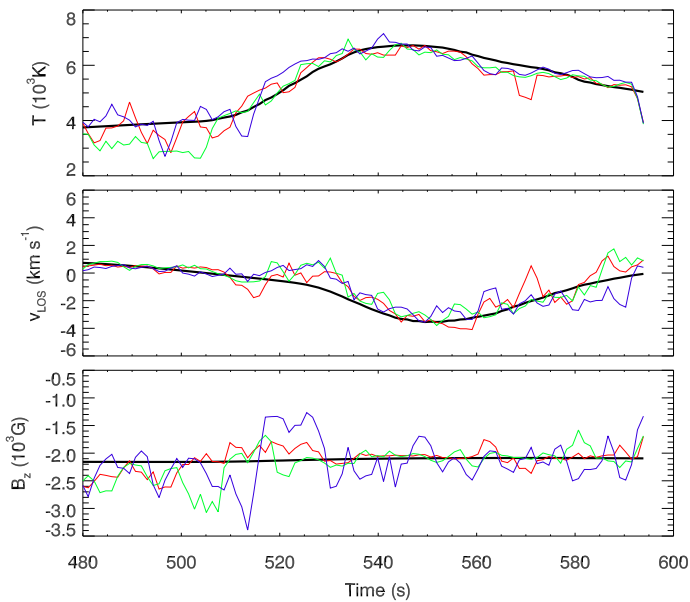}
  \caption{Comparison between temperature (top panel), LOS velocity (middle panel), and vertical velocity (bottom panel) at the chromosphere during a UF in the numerical simulation (black thick lines) and those inferred from the inversion of the synthetic Stokes profiles (color thin lines). The inversions were performed using different wavelength samplings. They correspond to cases \#3, \#7, and \#12 from the second column of Fig. \ref{fig:sampling} and share the same color code. They differ with regard to the resolution around the core of the line: $\delta\lambda=55$ m\AA\ (red line),  $\delta\lambda=80$ m\AA\ (green line), and $\delta\lambda=110$ m\AA\ (blue line).}    
  \label{fig:comparison_time}
\end{figure}

We focused on the comparison between the actual evolution of the temperature, LOS velocity, and vertical magnetic field at the chromosphere obtained from the numerical simulation during the UF and those values inferred from the inversion of the synthetic Stokes profiles. As the inversion is only sensitive to length scales that are long enough to modify the emerging radiation, the comparison was performed on the average atmosphere over a certain range of chromospheric optical depths. The magnetic field and temperature were averaged between $\log\tau=-4.6$ and $\log\tau=-5.1$, whereas the LOS velocity was averaged between $\log\tau=-4.6$ and $\log\tau=-4.9$. In the case of the latter, the results of the inversion were averaged at a slightly lower layer (between $\log\tau=-4.0$ and $\log\tau=-4.4$) than the simulated atmosphere since the inversions provide a better characterization of the chromospheric velocity at those layers. \citet{QuinteroNoda+etal2016} show that \CaII\ 8542 \AA\ line is highly sensitive to the region between $\log\tau=-4.5$ and $\log\tau=-5.5$, whereas \citet{Joshi+delaCruzRodriguez2018} find that the maximum response of this line to the magnetic field is between $\log\tau=-4.6$ and $\log\tau=-5.2$. 
 
The analysis was performed independently for all the set of wavelength samplings shown in Fig. \ref{fig:sampling}. Figure \ref{fig:comparison_time} shows the comparison between the temporal evolution of the simulated chromosphere and the temporal evolution inferred from the inversion of the synthetic Stokes profiles using a few wavelength samplings selected at random. The temporal variation of the inversions was slightly smoothed by applying a mean filter, replacing each value with the average of its two neighbors and itself. We note that the construction of the Stokes profiles requires a certain temporal span depending on the number of wavelength points and accumulations employed. For the results of the inversions, the time indicated at the horizontal axis represents the scanning time of the center of the \CaII\ 8542 \AA\ line, at the wavelength position $\Delta\lambda=0$.

We chose to illustrate three cases where the wavelength points were placed following criterion B, that is, using a constant resolution around the core of the line and setting an additional point at each wing around $\Delta\lambda=\pm0.9$ \AA. These examples correspond to cases with a fine resolution in the core of the line ($\delta\lambda=55$ m\AA, red line), medium resolution ($\delta\lambda=80$ m\AA, green line), and poor resolution ($\delta\lambda=110$ m\AA, blue line). The three inversions recover the general trends in the temperature and LOS velocity variations. 

The simulated LOS velocity exhibits a change from a slight downflow (positive velocity) to a strong upflow (negative velocity) when the shock wave reaches the chromosphere. At the beginning of the shock wave (around $t=515$ s), the case with finer resolution (red line) departs from the real value, first with a more negative velocity and then with a more positive velocity. This mismatch is an artifact attributed to the time required to complete the scanning of the line. When the UF starts to develop, the intensity of the core of the line increases as the core reversal is taking place. Since the line profile is scanned from the blue to the red, and each wavelength is acquired at a different time step, a spurious blueshift in the intensity minimum appears just prior to the UF and a redshift in the emission peak is found during the flash \citep{Felipe+etal2018a}. These deformations of the Stokes $I$ profile due to the time-dependent acquisition of various wavelengths are misinterpreted by the inversion code. Interestingly, the spurious blueshift at $t=515$ s is not present in the two cases with lower resolution (green and blue lines). Since these cases require less time to complete the scanning of the line (as they include a lower number of wavelength points), the spurious deformation of the Stokes profiles is reduced.   

The vertical magnetic field inferred from the inversions is around the value of 2100 G showed by the simulation at the chromosphere. However, significant fluctuations appear in the inverted results. These fluctuations increase with the reduction of the resolution. The case with lowest resolution (blue line) exhibits departures of the magnetic field from the actual value above 1000 G. The case with finer resolution (red line) shows a perfect agreement during a significant part of the illustrated temporal series, although mismatches of several hundred Gauss are also present.

\begin{figure}[!ht] 
 \centering
 \includegraphics[width=9cm]{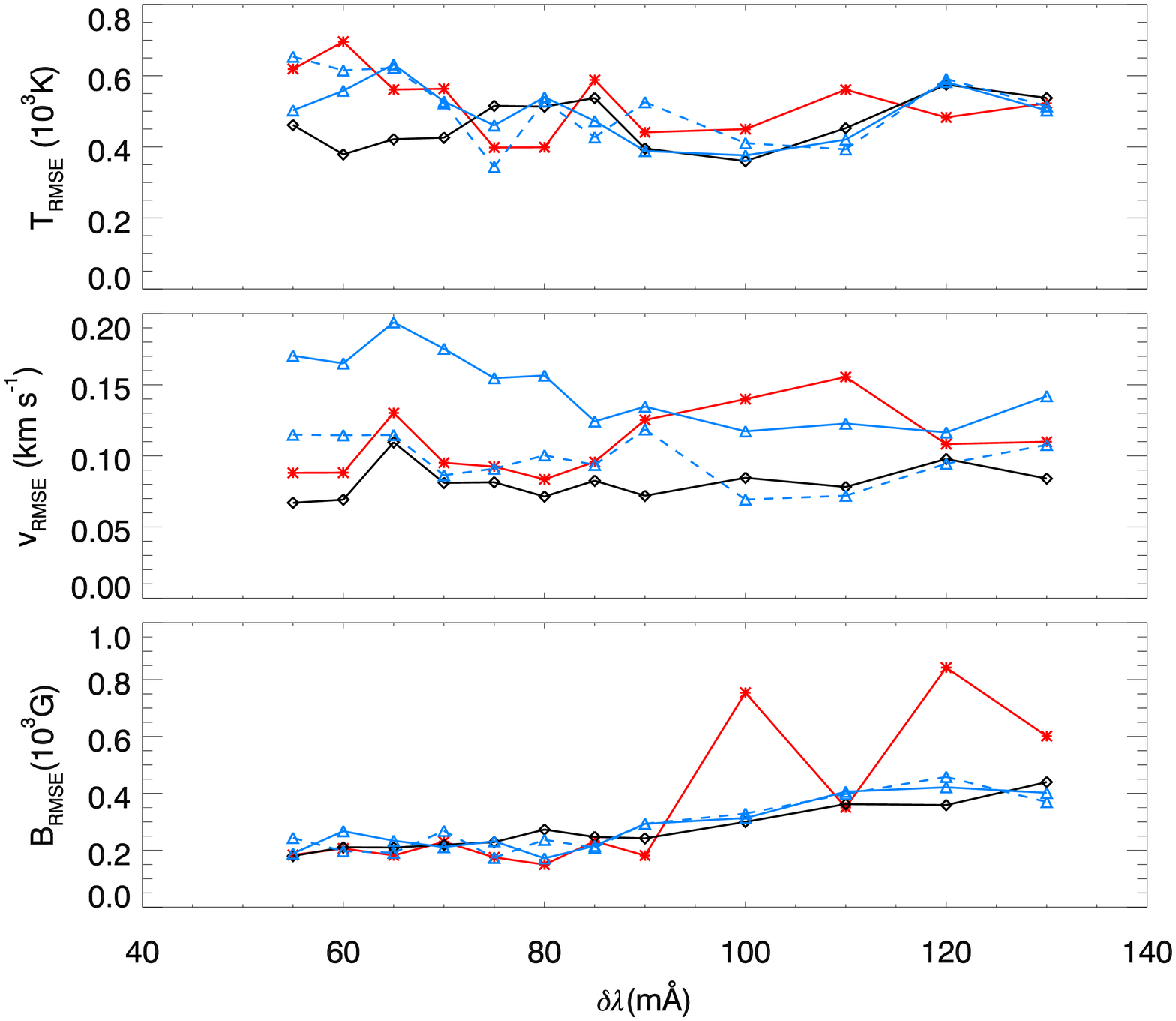}
  \caption{RMSE between inferences from the inversions and the actual values from the numerical simulation as a function of the wavelength resolution $\delta\lambda$ for the temperature (top panel), LOS velocity (middle panel), and vertical magnetic field (bottom panel). Each color line corresponds to a wavelength sampling selected following a different criteria. Blue line: criterion A (first column from Fig. \ref{fig:sampling}), black line: criterion B (second column from Fig. \ref{fig:sampling}), red line: criterion C (third column from Fig. \ref{fig:sampling}). The solid line shows the results for synthetic Stokes profiles constructed mimicking the acquisition strategy of imaging spectrographs (sequentially scanning the wavelengths of the profiles at successive time steps), whereas the dashed lines (only for criterion A) illustrate the error for synthetic profiles constructed with all the wavelength gathered at the same time step.}    
  \label{fig:RMSE}
\end{figure}

In order to evaluate the quality of the inversions retrieved from each of the wavelength samplings analyzed, the root-mean-square error (RMSE) between the inferred values at the chromosphere and the actual chromospheric values from the numerical simulation has been computed following

\begin{equation}
\theta_{\rm RMSE}=\sqrt{\frac{\sum_{j=1}^{j=n}(\theta_j^{\rm i}-\theta_j^{\rm s})^2}{n}},
\label{eq:RMSE}
\end{equation}

\noindent where $\theta$ is the evaluated variable (T for temperature, v for LOS velocity, and B for vertical magnetic field), $n=128$ is the number of time steps (from $t=445.0$ s to $t=597.4$ with a cadence of $1.2$ s), $\theta_j^{\rm i}$ is the chromospheric value of the variable $\theta$ from the inversion at time step $j$, and $\theta_j^{\rm s}$ is the chromospheric value of $\theta$ from the simulation at time step $j$. 

Figure \ref{fig:RMSE} shows the dependence of the RMSE for the temperature, LOS velocity, and vertical magnetic field with the resolution of the wavelength sampling. The color of the line indicates the criterion followed to select the position of the wavelength points, as described in Sect. \ref{sect:sampling}. The horizontal axis shows the resolution employed in each case around the core of the line.

The cases in which a uniform sampling was employed across the whole profile of the \CaII\ 8542 \AA\ line (criterion A, blue lines in Fig. \ref{fig:RMSE}) demonstrate some surprising results in the chromospheric temperature and velocity obtained from the inversion. The wavelength samplings with a fine resolution exhibit a poorer performance than those with a lower resolution. For the temperature, the best match between the inferred data and the actual values from the simulation is found for $\delta\lambda=100$ m\AA, whereas the velocity shows an approximately constant decrease of the RMSE with the reduction of the resolution. On the contrary, the vertical magnetic field shows the lowest RMSE for wavelength samplings with $\delta\lambda\le85$ m\AA. Above that limit, the RMSE progressively increases with the size of the wavelength step. The mismatches between the inferences from the inversion and the numerical simulation are partially due to the temporal span required by imaging spectrographs to complete the scan of the whole profile of the line. The dashed blue line in Fig. \ref{fig:RMSE} illustrates the RMSE for the instantaneous profiles, that is, those profiles that were constructed using the same temporal step of the numerical simulation for all the wavelengths. In this case, the inferred LOS velocity shows significantly lower differences with the actual values from the simulation.    

Criterion B uses a constant sampling around the core and adds two additional points and the blue and red wings, respectively (second column from Fig. \ref{fig:sampling}, black lines in Fig. \ref{fig:RMSE}). Profiles constructed following this method show the best performance for the determination of the chromospheric velocity, independently of the $\delta\lambda$. The RMSE of the velocity exhibits a slight positive trend, increasing for larger values of $\delta\lambda$. They also offer the best results in the chromospheric temperature for most values of $\delta\lambda$, including the cases with $\delta\lambda\le70$ m\AA\ and around $\delta\lambda\approx100$ m\AA. With regards to the magnetic field, their behavior is similar to that of the uniform sampling, showing a better agreement in the cases with finer resolution.

The red lines from Fig. \ref{fig:RMSE} illustrate the RMSE for the set of wavelength samplings constructed following criterion C (third column from Fig. \ref{fig:sampling}). In this case, the best matches for velocity are found for $\delta\lambda\le85$ m\AA\ (except at $\delta\lambda=65$ m\AA), whereas the agreement between the velocity inferences from the inversion and the actual chromospheric velocity from the simulation gets worse for low spectral resolutions. Similar results are found for the magnetic field, although for this variable, agreement is found for fine-resolution cases, up to $\delta\lambda=85$ m\AA.

\begin{figure}[!ht] 
 \centering
 \includegraphics[width=9cm]{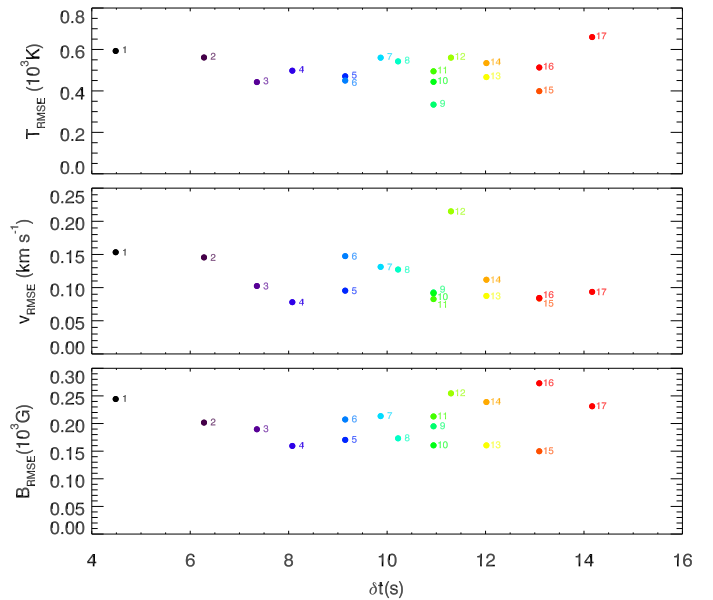}
  \caption{RMSE between the inferences from inversions and actual values from the numerical simulation as a function of the scanning time for temperature (top panel), LOS velocity (middle panel), and vertical magnetic field (bottom panel). The spectral samplings were selected following criterion D. Color code and numbers indicate the wavelength sampling employed in each case, as given in the fourth column of Fig. \ref{fig:sampling}. }    
  \label{fig:RMSE_dl80}
\end{figure}

Figure \ref{fig:RMSE_dl80} illustrates the performance of various wavelength samplings with the common factor of employing the same resolution of $\delta\lambda=80$ m\AA\ around the core of the \CaII\ 8542 \AA\ line (criterion D). We chose this $\delta\lambda$ based on the results plotted in Fig. \ref{fig:RMSE}  (and discussed above) since this resolution is fine enough to provide a good performance in the inference of the magnetic field, whereas its agreement for the temperature and LOS velocity is comparable to that of the values of $\delta\lambda$ that give the best results. A wavelength step of $\delta\lambda=80$ m\AA\ can offer a good compromise between the quality of the inversions and the time required to complete the scanning of the line.

In Fig. \ref{fig:RMSE_dl80}, the RMSE of the chromospheric temperature, LOS velocity, and vertical magnetic field is plotted as a function of the time required to complete the scanning of the spectral line. The wavelength samplings are identified by color and number, as defined in the fourth column of Fig. \ref{fig:sampling}. The RMSE measured for the fastest samplings (\#1, \#2, and \#3) shows that the region of the core covered in these samplings (wavelengths below $\Delta\lambda=\pm240$ m\AA) is insufficient for characterizing the umbral chromosphere. This is clearly seen in the fourth column of Fig. \ref{fig:sampling}, which shows that the wavelength sampling of the first three cases barely covers the intensity emission at the core of the line and the Stokes $V$ signal produced by the umbral flash. Scanning the region between $\Delta\lambda=\pm320$ m\AA\ (\#4, uniform sampling in that region) is sufficient to capture the chromospheric atmosphere of the simulated umbral flash. This wavelength sampling provides one of the best performances for inferring the chromospheric magnetic field and LOS velocity, together with sampling \#15 (uniform $\delta\lambda=80$ m\AA\ between $\Delta\lambda=\pm480$ m\AA\ and two points at each wing at $\Delta\lambda=\pm640$ m\AA\ and $\Delta\lambda=\pm800$ m\AA). The rest of the wavelength samplings explored show that adding more wavelength points towards the wings of the line (which requires a longer scanning time) does not necessary improve the results from the inversion at the chromosphere.

\section{Discussion and Conclusions}
\label{sect:conclusions}

Using synthetic data from the development of a numerical simulation of nonlinear wave propagation in a sunspot umbra, we evaluated the quality of NLTE inversions at the chromosphere during an UF. The basic approach taken in this work consists of (1) synthesizing the Stokes profiles from the temporal evolution of the simulated atmosphere; (2) constructing profiles comparable to those regularly observed, taking into account the noise of the data, the selected wavelength points, and scanning time across the profile of the \CaII\ 8542 \AA\ line; (3) inverting the synthetic profiles; and finally (4) comparing the properties inferred at the chromosphere with the actual values found in the numerical simulations. In this paper, we focus on the development of a parametric study of the quality of the inversion for different selections of the wavelength points included in the sampling of the line, which is crucial when observers are interested in studying highly dynamic events in the solar atmosphere with Fabry-P\'erot-based instruments. The detection of the event dynamics implies a compromise must be reached between the scanning time (proportional to the number of spectral points) and spectral resolution. The advent of a new generation of 4-m solar telescopes, such as DKIST and EST, will greatly mitigate the current limitations by using faster low-noise detectors and scanning instruments capable of providing high signal-to-noise spectral scans in a comparably shorter time span. In addition, next-generation instruments will perform integral field spectropolarimetry, allowing for simultaneous observations in both bidimensional spatial and spectral domains by employing several technical schemes, such as multi-slit image slicers or microlens arrays \citep{Jurcak+etal2019}. Integral field units have already been employed for the analysis of UFs \citep{Anan+etal2019}. Nevertheless, interferometer-based instruments will still play a key role in current and future solar telescopes.

Our results can be summarized as the following:

\begin{itemize}
\item The inferred vertical magnetic field strength is more precise when employing a fine wavelength resolution. The RMSE measured for the magnetic field increases with the length of the wavelength step, although no strong differences are found for wavelength samplings with $\delta\lambda\lessapprox 85$ m\AA. Even in the case of wavelength samplings that give the best results for the magnetic field during the UF, the error is about 200 G. This value is comparable to the magnetic field fluctuations attributed to UFs as reported by several authors \citep{Joshi+delaCruzRodriguez2018, Houston+etal2018}.

\item Using a proper wavelength sampling, one can infer the chromospheric temperature and LOS velocity associated to UFs with a RMSE below 500 K and 90 m s$^{-1}$, respectively. Previous observational works have measured temperature fluctuations around 1000 K and changes in the velocity up to 10 km s$^{-1}$ \citep[\eg,][]{SocasNavarro+etal2000b,delaCruz-Rodriguez+etal2013, Henriques+etal2017, Joshi+delaCruzRodriguez2018}.

\item During UFs, sudden changes are produced in the umbral chromosphere. These changes are significant enough to leave an imprint in the profiles scanned by imaging spectrographs since they require a certain time (generally above 10 s) to complete the record of narrow-band images at the selected wavelength points. Those spurious deformations of the profiles can be misinterpreted by the inversion code \citep{Felipe+etal2018a}. The study of rapidly-changing phenomena like UFs requires the use of acquisition strategies that allow a high temporal cadence. Fine wavelength resolution samplings can be counterproductive as they require a longer time to complete the scan of the whole profile. This is clearly illustrated in the cases where we employed an uniform wavelength step across the whole profile. In that set of inversions, cases with poorer resolution provide a better estimation of the LOS velocity than those with a fine resolution. Using fine resolution at the core of the line and lower resolution at the wings is generally considered a good strategy. 

\item The addition of wavelength points from the wings of the line (beyond $\Delta\lambda\approx\pm320$ m\AA\ in the case of the simulated UF) does not improve the agreement of the atmosphere inferred from the inversion with the actual values from the simulation at chromospheric layers. One difficulty in the inversion of spectropolarimetric data from the \CaII\ 8542 \AA\ using current Fabry-P\'erot instruments, such as IBIS or CRISP, is that the line is too broad to sample at least one point in the continuum. This is mainly due to the limited bandpass of the prefilter. Even though the wavelength positions at the wings are not necessary to infer the chromosphere, it may be convenient to include in the sampling the wavelength at a longer distance from the line center allowed by the prefilter in order to use it as pseudo-continuum.

\item The selection of the wavelength points must depend on the goal of the observations. Those observations focusing on the analysis of magnetic field fluctuations attributed to UFs should prioritize the wavelength resolution, whereas those focused on velocity and temperature could relax the spectral resolution and benefit from a better temporal cadence.  

\end{itemize}

Our analysis was performed for a single UF event. UFs produced by shocks with different properties (e.g.,\ with differences in the velocity amplitude or temperature fluctuations) would generate Stokes profiles whose characteristic features (emission core, strength of the lobes in Stokes $V$ and distance between them) are located at different wavelength positions. In such cases, differences may arise with regard to the ability of each of the explored wavelength samplings to capture the line profiles. In addition, the configuration of the inversion setup could probably be optimized individually for each distribution of wavelength points. For the purposes of consistency in this work, we employed the same inversion strategy for all the cases, including the same number of cycles, number and location of nodes, and the initial guess atmosphere. Finally, it must be noted that we only evaluated the quality of the inversions at the chromospheric layers. Observers who would be interested in extracting information from the \CaII\ 8542 \AA\ line at the photosphere should include more wavelength points at the wings of the line rather than the wavelength distribution taken from the cases favored in our analysis.

\begin{acknowledgements} 
Financial support from the State Research Agency (AEI) of the Spanish Ministry of Science, Innovation and Universities (MCIU) and the European Regional Development Fund (FEDER) under grant with reference PGC2018-097611-A-I00 is gratefully acknowledged. SEP acknowledges the financial support received from the European Research Council (ERC) under the European Union's Horizon 2020 research and innovation programme (POLMAG, Grant agreement No 742265). The authors wish to acknowledge the contribution of Teide High-Performance Computing facilities to the results of this research. TeideHPC facilities are provided by the Instituto Tecnol\'ogico y de \hbox{Energ\'ias} Renovables (ITER, SA). URL: http://teidehpc.iter.es. 
\end{acknowledgements}

\bibliographystyle{aa} 
\bibliography{biblio.bib}

\end{document}